\documentclass[aps,preprint,a4paper,showpacs,showkeys,superscriptaddress]{revtex4-1}

\ifx\pdfoutput\undefined
\usepackage[usenames]{color} 
\usepackage[dvips]{graphicx}
\else
\usepackage{color} 
\definecolor{BrickRed}{rgb}{0.85,0.15,0.25}
\definecolor{MidnightBlue}{rgb}{0,0.45,0.85}
\definecolor{ForestGreen}{rgb}{0,0.85,0.45}
\usepackage{graphicx}
\usepackage{epstopdf}
\fi
\usepackage{latexsym,amsmath,amssymb}
\allowdisplaybreaks

\usepackage{umoline}
\newsavebox\CBox

\begin{document}


\title{Freely Falling Observer and Black Hole Radiation}

\author{Wontae Kim}
\email[]{wtkim@sogang.ac.kr}
\affiliation{Department of Physics, Sogang University, Seoul 121-742, Republic of Korea}
\affiliation{Center for Quantum Spacetime, Sogang University, Seoul 121-742, Republic of Korea}
\affiliation{Research Institute for Basic Science, Sogang University, Seoul, 121-742, Republic of Korea}

\author{Edwin J. Son}
\email[]{eddy@nims.re.kr}
\affiliation{Division of Computational Sciences in Mathematics,
National Institute for Mathematical Sciences, Daejeon 305-811, Republic of Korea}

\date{\today}

\begin{abstract}
We find radiation in an infalling frame and 
present an explicit analytic evidence of the failure of 
no drama condition by showing that an infalling observer finds 
an infinite negative energy density at the event horizon.
The negative and positive energy density regions
are divided by the newly defined zero-energy curve. 
The evaporating black hole is surrounded by the negative energy
which can also be observed in the infalling frame.
\end{abstract}



\maketitle


Since the discovery of Hawking's thermal radiation from a black hole has raised an intriguing information loss problem~\cite{Hawking:1974rv,Hawking:1974sw,Hawking:1976ra}, there have been intensive studies  in
the quantum-gravity arena. In particular, it has been proposed that 
the information can be significantly released after the Page time~\cite{Page:1993df,Page:1993wv}. 
On the other hand,
the information cloning problem can also be overcome by black hole complementarity
(BHC)~\cite{Susskind:1993if,Susskind:1993mu,Stephens:1993an}, which claims that there is no contradictory
 physical process between the freely falling observer and the distant observer. Recently, Almheiri, Marolf, Polchinski, 
and Sully (AMPS)~\cite{Almheiri:2012rt} have suggested an amazing puzzle referred to as the firewall paradox of quantum black holes (for a similar prediction from different assumptions, see~\cite{Braunstein:2009my}).
 They argued that a freely falling observer experiences something special near the horizon and burns up 
 because of high energy quanta.  Subsequently, much attention
has been paid to resolve this problem along with the information loss problem from various viewpoints~\cite{Bousso:2012as,Nomura:2012sw,Susskind:2012rm,Banks:2012nn,Ori:2012jx,Susskind:2012uw,Hossenfelder:2012mr,Page:2012zc,Giddings:2012gc,Jacobson:2012gh,Kim:2013fv,Giddings:2013kcj,Almheiri:2013hfa,Verlinde:2013uja,Maldacena:2013xja,Giddings:2013vda,Almheiri:2013wka}. 

Now, we are going to investigate whether radiation can be found in an infalling frame 
or not by using the amenable setting 
called Callan-Giddings-Harvey-Strominger (CGHS)
model~\cite{Callan:1992rs}, which is consistent, renormalizable, and exactly soluble classically. Moreover, the Hawking flux can be exactly  calculated semiclassically in an evaporating black hole. 
First of all, we assume simply two things: one is that energy-momentum tensors transform as true tensors without any anomalies, and the other is that the semiclassical equations of motion are valid. We do not have to postulate the complete evaporation of the
black hole which plays an important role in the
firewall argument~\cite{Almheiri:2012rt}.  It implies that the present argument 
has nothing to do with the remnant issue in~\cite{Almheiri:2013wka,Ashtekar:2008jd}.
On the other hand, it has also been claimed that there is no apparent need for firewalls because unitary evolution of black hole entangles a late mode located outside the horizon with a combination of early radiation and black hole states, instead of either of them separately~\cite{Hutchinson:2013kka}.
The aim of the present work will be to show that there can exist non-trivial effect at the horizon based on the conventional quantum field theory without resort to the firewall argument.

In this work, we will show that there exists radiation
in the infalling frame, in particular, the infinite negative energy density at the horizon,
which is related to the failure of no drama condition at the event horizon.
Moreover, we introduce a newly defined zero-energy curve (ZEC) dividing spacetime into the negative energy region 
and the positive energy region.

Let us start with the two-dimensional dilaton gravity coupled to massless scalar fields
given by CGHS~\cite{Callan:1992rs}
\begin{equation}
S = \frac{1}{2\pi} \int d^2x \sqrt{-g} \left[ e^{-2\phi} \left( R + 4(\nabla \phi)^2 + 4\lambda^2 \right) - \frac12 \sum_{i=1}^{N} (\nabla f_i)^2 \right],
\end{equation}
where $\phi$ is a dilaton field, $f_i$ are scalar fields, and $N$ is the number of scalar fields.
For the conformal gauge of $ds^2 = -e^{2\rho} dx^+ dx^-$ in the light-cone coordinates,
the equations of motion and constraints can be solved 
for the shock wave described by the energy-momentum tensor 
$ (1/2) \sum_i \partial_+ f_i \partial_+ f_i = M \delta(x^+ - x^+_0) $.
Thus, one can find the solution 
$e^{-2\rho} = e^{-2\phi} = - M(x^+ - x^+_0) \Theta (x^+ - x^+_0) - \lambda^2 x^+ x^- $ in the Kruskal coordinates. Next, one can take 
the coordinate transformation as 
$e^{\lambda \sigma^+} = \lambda x^+ $ and $ e^{-\lambda \sigma^-} = - \lambda x^- - 
M/\lambda$.
Then, the metric can be written in the form of  
\begin{equation}
\label{metric}
e^{2\rho} = \left\{
\begin{aligned}
& \left[ 1+ \frac{M}{\lambda} e^{\lambda \sigma^-} \right]^{-1} && \text{for}\ \sigma^+ < \sigma^+_0, \\
& \left[ 1+ \frac{M}{\lambda} e^{-\lambda (\sigma^+ - \sigma^- - \sigma^+_0)} \right]^{-1} && \text{for}\ \sigma^+ > \sigma^+_0.
\end{aligned}
\right.
\end{equation}

As for the Hawking radiation~\cite{Callan:1992rs}, one can use the one-loop trace anomaly of 
$\langle T^f_{+-} \rangle = - \kappa \partial_+ \partial_- \rho$ with $\kappa=N/12$,
while the covariant conservation of the energy-momentum tensors is maintained, 
which yields the energy-momentum tensors as
\begin{equation}
\label{eff:stress}
\langle T^f_{\pm\pm} \rangle = - \kappa \left[ (\partial_\pm \rho)^2 - \partial_\pm^2 \rho + t_\pm (\sigma^\pm) \right],
\end{equation}
where the functions $t_\pm$ reflect the nonlocality of the trace anomaly. 

Note that the CGHS model has some important properties of the event horizon and the curvature singularity like the four-dimensional back holes, and especially the conformal anomaly for the scalar fields can be employed to calculate the energy-momentum tensors. On general grounds, however, one may consider a realistic scalar field on the Schwarzschild black hole background, then it will be a non-trivial task to realize the conformal anomaly or the effective action for the matter field directly from  dimensional reduction at the quantum level. If we consider a four-dimensional scalar field on the background of the Schwarzschild black hole, the original four-dimensional action for 
the scalar field can be represented as a sum over modes of two-dimensional effective action before renormalization; however, this is not the case generically after renormalization because of dimensional-reduction anomaly~\cite{Frolov:1999an}. This comes from the fact that the four-dimensional renormalization is not equivalent to renormalization of the two-dimensional effective theory since the number of divergent terms depends on the number of dimensions. So the modified effective action may give a different type of energy-momentum tensors which are related to Hawking radiation. In the above, we simply took the conventional trace anomaly for two-dimensional scalar fields without taking into account the origin of the scalar fields for simple argument.

\begin{figure}[pt]
  \begin{center}
  \includegraphics[width=0.65\textwidth]{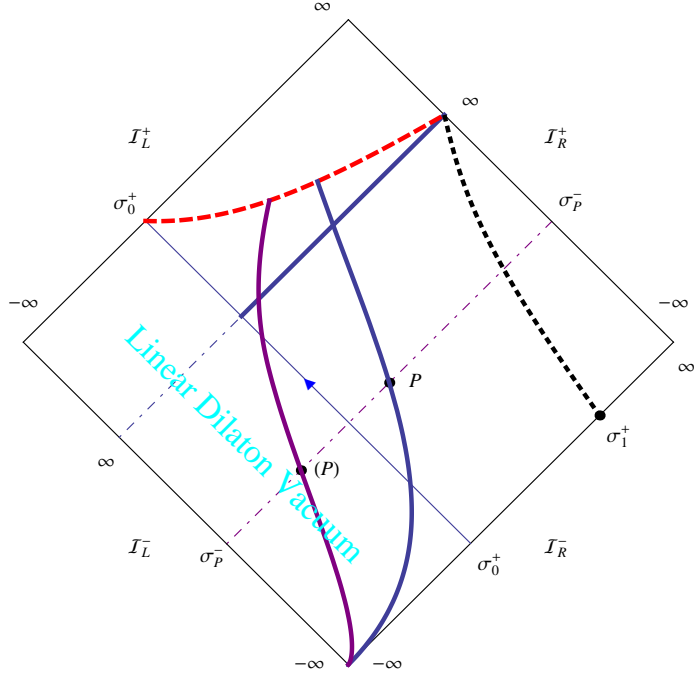}
  \end{center}
  \caption{An incoming shock wave collapses to a black hole. The spacelike curve represents the curvature singularity and the event horizon is located at $\sigma^- = \infty$. The two lower timelike curves describe the geodesics for infalling observers. The uppermost timelike curve represents ZEC starting from $\sigma^+_1 = \sigma^+_0 + 2\lambda^{-1} \ln 2$. }
  \label{fig:penrose} 
\end{figure}

Before we get down to the calculations of the radiations, let us define a coordinate transformation from a fixed observer ($\sigma^{\pm}$) 
to a freely falling observer ($\tilde{\sigma}^{\pm}$) satisfying the geodesic equations given by
\begin{equation}
\label{eq:geodesic}
0=\left( \frac{d^2 \tilde{\sigma}^{\pm}}{d \tau^2} \right)_{P} = \left( \frac{\partial \tilde{\sigma}^{\pm}}{\partial \sigma^{\pm}} \right) \left[ \frac{d^2 \sigma^{\pm}}{d \tau^2} + \Gamma^\pm_{\pm\pm} \frac{d \sigma^{\pm}}{d \tau} \frac{d \sigma^{\pm}}{d \tau}\right]_{P}
\end{equation}
at a particular point $P$, where the affine connections are defined by 
$\Gamma^\pm_{\pm\pm} |_{P} \equiv (\partial_\pm \tilde{\sigma}^{\pm})^{-1} (\partial_\pm^2 \tilde{\sigma}^{\pm}) |_{P}$. 
Then, the transformation from the fixed coordinates to the locally flat coordinates in the vicinity of the point $P$ up to  second order is implemented by
\begin{equation} 
\label{local:coord}
\tilde{\sigma}^{\pm} 
= \tilde{\sigma}^\pm_P + b^\pm_{~\pm} (\sigma^\pm - \sigma^\pm_P) + \frac12 b^\pm_{~\pm} \Gamma^\pm_{\pm\pm}|_{P} (\sigma^\pm - \sigma^\pm_P)^2,
\end{equation}
where $\tilde{\sigma}^\pm_P = \tilde{\sigma}^{\pm} (\sigma^\pm_P)$ are arbitrary constants.
Actually, the point $P$ can be located either in the linear dilaton-vacuum region or in the 
black hole region as seen in Fig.~\ref{fig:penrose}.
From the metric~\eqref{metric}, the explicit form of 
$ b^\pm_{~\pm}$ can be calculated as  
\begin{equation}
  \label{transf:factor}
  b^\pm_{~\pm} = \left. \frac{\partial \tilde{\sigma}^\pm}{\partial \sigma^\pm} \right|_{P}
  = \left[ 1 + \frac{M}{\lambda} e^{-\lambda((\sigma^+_P - \sigma^+_0) \Theta (\sigma^+_P - \sigma^+_0) - \sigma^-_P)} \right]^{-1/2},
\end{equation}
which makes the metric 
become the local Minkowski spacetime at the point $P$, {\it  i.e.}, 
$e^{-2\tilde{\rho}(\tilde{\sigma}^+_P,\tilde{\sigma}^-_P)}=1$.
According to the transformation~\eqref{local:coord},
the metric in the freely falling frame around the point $P$ can also be written as $ds^2= -e^{2\tilde{\rho}(\tilde{\sigma}^+,\tilde{\sigma}^-)} d\tilde{\sigma}^+ d\tilde{\sigma}^-$
where 
\begin{equation}
\label{local:metric}
\begin{aligned}
e^{-2\tilde{\rho}(\tilde{\sigma}^+,\tilde{\sigma}^-)} = b^+_{~+} b^-_{~-} e^{-2\rho(\sigma^+,\sigma^-)} \Big[ 1 &+ \Gamma^+_{++} |_{P} (\sigma^+ - \sigma^+_P) + \Gamma^-_{--} |_{P} (\sigma^- - \sigma^-_P) \\
  & + \Gamma^+_{++}|_{P} \Gamma^-_{--} |_{P} (\sigma^+ - \sigma^+_P) (\sigma^- - \sigma^-_P) \Big].
\end{aligned}
\end{equation}
Note that the freely falling frame was defined at each point so that the frame 
is dropped from the point $P$ with  zero velocity. 
For a nonvanishing velocity, Eq.~\eqref{transf:factor} should be modified in such a way 
that it contains a velocity-parameter as shown in Ref.~\cite{Ford:1993bw}. 
Now, we are in a position to study what happens when the infalling 
observer sees nothing special. 

Now,
we will assume that there exists 
no radiation in the linear dilaton-vacuum and then 
investigate whether the radiation exists or not in the infalling frame. 
For this purpose, it is required 
that in the linear dilaton region, ${}_\text{in}\langle T^f_{\pm\pm} (\sigma^+ < \sigma^+_0,\sigma^-) \rangle_\text{in} = 0$, 
where the vacuum is denoted by
the in-vacuum $|0\rangle_{\text{in}}$, 
which determines the boundary functions $t_+(\sigma^+) = 0$ and $t_-(\sigma^-) = -(\lambda^2/4) [ 1 - ( 1 + (M/\lambda) e^{\lambda \sigma^-} )^{-2} ]$~\cite{Bilal:1992kv,Kim:1995wr}. 
So the energy-momentum tensor on the line $\sigma^- = \sigma^-_P$ can be calculated as 
\begin{equation}
\label{rad:in:bulk}
\begin{aligned}
{}_\text{in}\langle T^f_{--} (\sigma^+ > \sigma^+_0,\sigma^-_P) \rangle_\text{in} &= \frac{\kappa\lambda^2}{4} \left[ \left( 1 + \frac{M}{\lambda} e^{-\lambda(\sigma^+ - \sigma^-_P - \sigma^+_0)} \right)^{-2} - \left( 1 + \frac{M}{\lambda} e^{\lambda \sigma^-_P} \right)^{-2} \right] \\
  &\xrightarrow{\sigma^+ \to \infty}  \frac{\kappa \lambda^2}{4} \bigg[ 1 - \left( 1 + \frac{M}{\lambda} e^{\lambda \sigma^-_P} \right)^{-2} \bigg].
\end{aligned}
\end{equation}
Note that the boundary functions correspond to the unknown constants of the auxiliary field to localize the quantum effective action~\cite{Mottola:2006ew}. 
These constants are determined by the boundary conditions and related to the divergent structure in the Schwarzschild black hole in two and four dimensions.

Next, the coordinate transformation to the infalling frame is performed as
\begin{align}
\label{fire}
  &{}_\text{in}\langle \tilde{T}^f_{--} (\tilde{\sigma}^+,\tilde{\sigma}^-) \rangle_\text{in} |_{\sigma^+ > \sigma^+_0,\sigma^-=\sigma^-_P} \notag \\
  &= (b^-_{~-})^{-2} {}_\text{in}\langle T^f_{--} (\sigma^+ > \sigma^+_0,\sigma^-_P) \rangle_\text{in} \notag \\
  & = \frac{\kappa\lambda^2}{4} \left( 1 + \frac{M}{\lambda} e^{-\lambda( \sigma^+_P - \sigma^-_P - \sigma^+_0)} \right) \times \notag \\
  & \quad \times \bigg[ \left( 1 + \frac{M}{\lambda} e^{-\lambda[\sigma^+_P + (b^+_{~+})^{-1} (\tilde{\sigma}^+ - \tilde{\sigma}^+_P) + O(\tilde{\sigma}^+ - \tilde{\sigma}^+_P)^2 - \sigma^-_P - \sigma^+_0]} \right)^{-2} - \left( 1 + \frac{M}{\lambda} e^{\lambda \sigma^-_P} \right)^{-2} \bigg],
\end{align}
where the transformation~\eqref{local:coord} was employed.
The nonvanishing outgoing radiation~\eqref{fire} in the infalling frame at the point $P$ 
can be found as
\begin{align}
\label{rad:ff:in}
{}_\text{in}\langle \tilde{T}^f_{--} (\tilde{\sigma}^+_P,\tilde{\sigma}^-_P) \rangle_\text{in} |_{\sigma^+_P > \sigma^+_0} = \frac{\kappa\lambda^2}{4} \bigg[ & \left( 1 + \frac{M}{\lambda} e^{-\lambda(\sigma^+_P - \sigma^-_P - \sigma^+_0)} \right)^{-1} \notag \\
  & - \left( 1 + \frac{M}{\lambda} e^{-\lambda(\sigma^+_P - \sigma^-_P - \sigma^+_0)} \right) \left( 1 + \frac{M}{\lambda} e^{\lambda \sigma^-_P} \right)^{-2} \bigg].
\end{align}
Note that the outgoing radiation is zero at the horizon which is a conventional result. By the way,
the radiation~\eqref{rad:ff:in} for $\sigma^+_P \to \infty$ tells us that it is interestingly coincident  with the Hawking flux~\eqref{rad:in:bulk}, so that the extreme infalling observer 
at the null infinity $\mathcal{I}_R^+$ can
detect the same amount of radiation with the Hawking radiation.  
It seems to be plausible in that it is impossible to distinguish physically the two 
asymptotic observers (for a realistic four-dimensional  numerical analysis, see~\cite{Greenwood:2008zg}.)

Now, the ingoing flux on the line $\sigma^+ = \sigma^+_P > \sigma^+_0$ measured by the fixed observer is given by
\begin{equation}
\label{tpp}
{}_\text{in}\langle T^f_{++} (\sigma^+_P > \sigma^+_0,\sigma^-) \rangle_\text{in}
= - \frac{\kappa \lambda^2}{4} \left[ 1 - \left( 1 + \frac{M}{\lambda} e^{-\lambda(\sigma^+_P - \sigma^- - \sigma^+_0)} \right)^{-2} \right],
\end{equation}
which is negative and finite at the horizon, \textit{i.e.}, $-\kappa \lambda^2 /4$.
By using the transformation~\eqref{local:coord}, the ingoing flux in the infalling frame
at the point $P$ is obtained as
\begin{equation}
\begin{aligned}
{}_\text{in}\langle \tilde{T}^f_{++} (\tilde{\sigma}^+_P,\tilde{\sigma}^-_P) \rangle_\text{in} 
= - \frac{\kappa\lambda^2}{4} \left[ \left( 1 + \frac{M}{\lambda} e^{-\lambda(\sigma^+_P - \sigma^-_P - \sigma^+_0)} \right) - \left( 1 + \frac{M}{\lambda} e^{-\lambda(\sigma^+_P - \sigma^-_P - \sigma^+_0)} \right)^{-1} \right],
\end{aligned}
\end{equation}
where the ingoing flux is negatively divergent at the horizon. 
It is interesting to note that this divergent flux does not mean that a physical observer will observe all particle states because the detector cannot register whose wavelength is much larger than the size of the detector. Actually, the size of the freely falling detector  based on the spirit of the local inertial frame should be taken as very small enough to smooth out the tidal force so that only the high frequency modes can be detected in the detector and the detector will miss most of the particles from the horizon. In this respect, an infalling observer might not burn at the horizon.

On the other hand, using Eqs.~\eqref{rad:in:bulk} and \eqref{tpp}, the energy density $\epsilon$  in the fixed coordinates 
is calculated as \begin{equation}
\epsilon = e^{-4\rho} \left[ \langle T^{f}_{++} \rangle + \langle T^{f}_{--} \rangle + 2 \langle T^{f}_{+-} \rangle \right],
\end{equation}
with the help of the trace anomaly.
Note that it  vanishes in the linear dilaton-vacuum region as expected,
while in the black hole region $\sigma^+ > \sigma^+_0$, it is explicitly written as
\begin{equation}
\epsilon = \frac{\kappa \lambda^2}{4} \bigg\{ 2 - \frac{4M}{\lambda} e^{-\lambda(\sigma^+ - \sigma^- - \sigma^+_0)} - \left( 1 + \frac{M}{\lambda} e^{-\lambda(\sigma^+ - \sigma^- - \sigma^+_0)} \right)^{2} \bigg[ 1 + \left( 1 + \frac{M}{\lambda} e^{\lambda \sigma^-} \right)^{-2} \bigg] \bigg\}.
\end{equation}
At the horizon, the energy density is negatively divergent, whereas it is positive finite far from the horizon. 
The energy-momentum tensors of $\langle T^{f}_{\pm\pm} \rangle$ and $\langle T^{f}_{+-} \rangle$ are regular everywhere~\cite{Fulling:1978ht}; however, in our case the divergence comes from the energy density $\epsilon=\langle T_{f}^{00} \rangle$.
So, we can naturally define ZEC to distinguish two regions 
by imposing condition of $\epsilon=0$, which gives a
curve starting from $\sigma^+_1$ shown in Fig.~\ref{fig:penrose}:
\begin{equation}
\label{str:hor}
e^{-\lambda (\sigma^+ - \sigma^+_0)} = \frac{1+\zeta}{1+3\zeta^2+\sqrt{2\zeta^2(3+5\zeta^2)}},
\end{equation}
where $\zeta = 1 + (M/\lambda) e^{\lambda \sigma^-}$.
If one considers a spacelike curve from the event horizon
to the asymptotic future null infinity,  
the energy density increases from the negative state at the event horizon 
to the positive Hawking radiation region across ZEC.
At last, the energy density in the infalling frame in the black hole region of $\sigma^+ > \sigma^+_0$ is calculated as
\begin{equation}
\begin{aligned}
\tilde{\epsilon} = \frac{\kappa \lambda^2}{4} \bigg\{ & 2 \left( 1 + \frac{M}{\lambda} e^{-\lambda(\sigma^+ - \sigma^- - \sigma^+_0)} \right)^{-1} - \left( 1 + \frac{M}{\lambda} e^{-\lambda(\sigma^+ - \sigma^- - \sigma^+_0)} \right) \bigg[ 1 + \left( 1 + \frac{M}{\lambda} e^{\lambda \sigma^-} \right)^{-2} \bigg]  \\
& - \frac{4M}{\lambda} e^{-\lambda(\sigma^+ - \sigma^- - \sigma^+_0)} \left( 1 + \frac{M}{\lambda} e^{-\lambda(\sigma^+ - \sigma^- - \sigma^+_0)} \right)^{-1} \bigg\}.
\end{aligned}
\end{equation}
The well-known Hawking radiation flux is exactly recovered at the future null infinity and
the energy density also vanishes on ZEC~\eqref{str:hor}. 
At the horizon, the infinite negative energy density appears in the infalling frame,
which implies the failure of the no drama condition. 


In conclusion, 
we have shown that the infalling observer can find the negative energy zone
around the black hole,
especially an infinite negative energy density at the event
horizon. We also discussed Hawking radiation  between
the infalling  observer and the distant observer.
Furthermore, it will be interesting to extend this work by taking into account the back reaction of the geometry, since there is no divergence at the horizon in the back reacted model in Ref.~\cite{Ashtekar:2010hx}.

The final comment to be mentioned is that the energy density at the horizon in the freely falling frame is divergent; however, it has been expected to be finite~\cite{Birrell:1982ix}, since the energy density due to radiation can be cancelled by the negative energy density of the vacuum polarization near the horizon. In our calculations, the divergent energy density was measured by the freely falling observer who starts to move just at the horizon without the long-term journey. If the free fall happens at a far distance from the horizon, then the energy density measured by the freely falling observer will  be finite because the energy flux due to the positive Hawking radiation can be cancelled out by sweeping out through the cloud of the vacuum polarization as was claimed in the standard argument~\cite{Birrell:1982ix}. If the back reaction of the geometry were taken into account together, the more rigorous investigation was possible. We hope this issue will be discussed elsewhere in more detail.

\acknowledgments 
We are grateful to E. Choi, M. Eune, Y. Gim, M. Kim, and S.-H. Yi for exciting discussions.
W. K. was supported by the Sogang University Research Grant of (2013)201310022.


\end{document}